\documentclass[aps,prl,superscriptaddress,twocolumn,tightenlines,showpacs]{revtex4}

\usepackage {graphicx}
\usepackage {amsmath}
\usepackage {amsfonts}
\usepackage {amssymb}
\usepackage {mathrsfs}

\usepackage{epstopdf}

\newcommand {\be}{\begin{eqnarray}}
\newcommand {\ee}{\end{eqnarray}}

\begin{document}

\title {The quantum smectic as a dislocation Higgs phase.}
\author {V.\ Cvetkovic}
\email {vladimir@lorentz.leidenuniv.nl}
\affiliation {Instituut Lorentz voor de theoretische natuurkunde, 
Universiteit Leiden, P.O. Box 9506, NL-2300 RA Leiden, 
The Netherlands}
\author {J\. Zaanen}
\email {jan@lorentz.leidenuniv.nl}
\affiliation {Instituut Lorentz voor de theoretische natuurkunde, 
Universiteit Leiden, P.O. Box 9506, NL-2300 RA Leiden, 
The Netherlands}

\date {\today}
\begin {abstract}
The theory describing quantum-smectics in 2+1
dimensions, based on
topological quantum melting is presented. This
is governed by a dislocation condensate characterized
by an ordering of Burger's vector and this `dual shear 
superconductor' manifests itself in the form of a
novel spectrum of phonon-like modes. 
\end {abstract}


\maketitle

Different from classical liquid crystals \cite {deGennes, ChaikinLubensky},
the quantum smectic and nematic type 
orders occurring at zero temperature are far from understood. These came into
focus recently, motivated  by empirical  developments in high
$T_c$ superconductivity 
and quantum Hall systems \cite {KFE}. Fundamentally, it is about the
partial breaking of the symmetries of space itself, and on the  quantum 
level this might carry consequences  which cannot be envisaged classically. 

The `most ordered' liquid crystal is the smectic,
which can be pictured as  lines in 2D (or layers in 3D) of liquid 
forming a periodic array in one spatial direction. 
Emery {\em et al.} \cite {KivelsonLubensky}
(for doped Mott insulators) and MacDonald and Fisher \cite {MacDonaldFisher}
(for quantum Hall systems), delivered proof of principle
that things are different on the quantum level by showing  
that a two dimensional quantum system can organize 
spontaneously into an array of one dimensional metals.     
Here we will present a description of the quantum smectic 
which is  complementary to these earlier works. It rests on Kramers-Wannier
duality \cite{RSavit, Kleinert1}, the
field-theoretical fact that the disordered state (the smectic) corresponds to an ordered
state (in fact, the Higgs phase) formed from the topological excitations (dislocations) of the
ordered state (the crystal), and as such it can be viewed as a quantum extension of
the famous Nelson-Halperin-Young \cite {NHY} theory of two dimensional melting.

\begin{figure*} 
\includegraphics[width=0.98\textwidth]{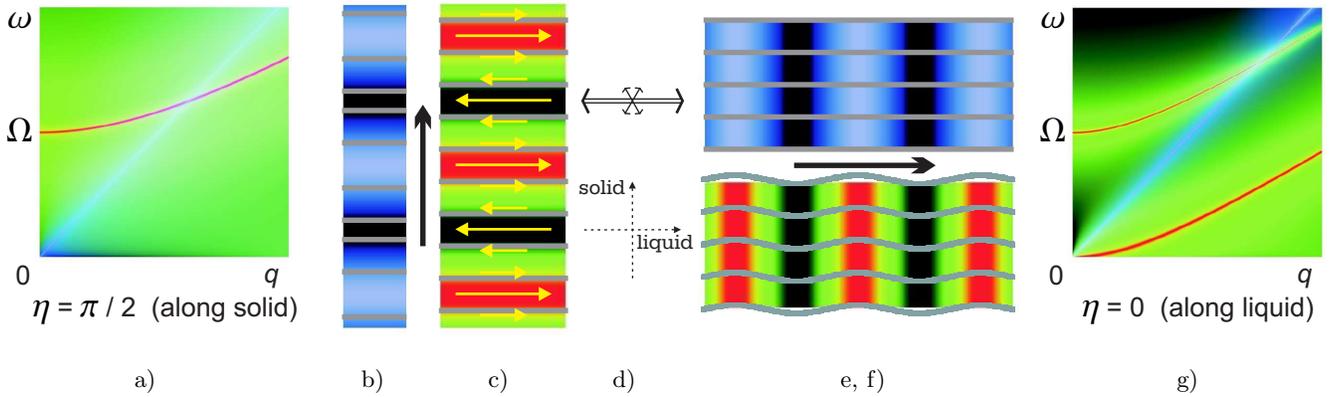}
\begin {center} \hspace* {0.0\textwidth} a) \hspace {0.14\textwidth} b)
\hspace {0.065\textwidth} c) \hspace {0.065\textwidth} d)
\hspace {0.14\textwidth} e, f)\hspace {0.21\textwidth} g)
\hspace* {0.0\textwidth} \end {center}
\caption{ (color). The elastic response (strain spectral functions)
of a quantum smectic as function of momentum ($q$) and 
frequency ($\omega$) in the two
special solid (a) and  liquid (g) direction, as determined 
by the orientation of the condensate Burgers vectors (d).
The longitudinal phonons (blue) are unaffected, corresponding
either to (b) compression of layers along the solid direction or
to (e) density waves in the liquid direction. The transversal
response (red) is gapped along the solid direction due to the
decoupling of the layer motion (c).  Along
the liquid direction (f), shear rigidity is screened,
but the transversal motions of the layers results in a
massless  quadratic mode.}
\label {FigBig}
\end{figure*}

The theory is completely tractable for a system of bosons 
living in the 2+1D galilean invariant 
continuum \cite{ZMN},  in the
limit that all characteristic length scales are large compared to the
lattice constant. The  outcome is a spectrum of propagating long-wavelength
collective modes \cite{thesis}, 
which should have  a universal status 
in the scaling limit. Before discussing the theory, let us first
present this mode spectrum.  The quantum smectic in 2+1D
is characterized by a `crystalline' and 
an orthogonal `liquid' direction (see Fig.\ \ref {FigBig}d).
For simplicity we assume that the quantum smectic is
associated with a reference crystal described by isotropic quantum elasticity (e.g.,
a hexagonal crystal) characterized by just a shear-($\mu$) and compression ($\kappa$)
modulus, and a mass density $\rho$, such that the 
longitudinal and transversal phonon velocities are given by 
$c_L = \sqrt{ ( \kappa + \mu )/\rho}$ and   $c_T = \sqrt{  \mu /\rho}$.
For modes propagating exactly along either the
fluid ($\eta = 0 \mod \pi$, i.e., Figs.\ \ref {FigBig}abc)
or solid ($\eta = \pi/2 \mod \pi$, i.e., Figs.\ \ref {FigBig}efg)
directions  transversal and longitudinal motions decouple.
Let us first focus on the solid direction.
The longitudinal mode (Fig.\ \ref {FigBig}b) vibrates the 
`leftover' lattice and behaves like a phonon; 
it actually propagate with $c_L$, showing that this mode is set by
the shear rigidity of the reference crystal. The transversal 
mode (Fig.\ \ref {FigBig}c), corresponding to `sliding motions'
of the `liquid lines' relative to each other, is very 
different from a phonon:  there is no massless mode and we find instead a single 
massive `dual Higgs photon' (Fig.~\ref {FigBig}a)
which is the fingerprint of the `true' liquid as we will explain later.
This implies the absence of a reactive elastic 
response, and our quantum smectic can be understood as 
`an array of independent one dimensional quantum liquids', in
the same sense as the  Kivelson-Lubensky 
sliding phases \cite {KivelsonLubensky}.

Let us now turn to the modes propagating along
the fluid direction (Fig.\ \ref {FigBig}g). 
The longitudinal mode (Fig.\ \ref {FigBig}e)
looks like a compressional wave in
the liquid, but it is again indistinguishable from the longitudinal phonon of
the reference crystal. The transversal wave in the  liquid direction 
(Fig.\ \ref {FigBig}f) is clearly not fluctuating the lattice directly, but it is 
neither a pure motion of  the liquid. The outcome is remarkable:
besides the `dual Higgs photon', the fingerprint of the liquid, we find a
mode with a quadratic dispersion: $\omega = \lambda_S c_T q^2$, where $\lambda_S$ 
is the shear penetration depth,  the characteristic
length of crystal correlations in the liquid, associated with
the `shear Higgs mass' by $\Omega = c_T / \lambda_S$.
These quadratic modes are of course well known from 
the classical smectics \cite{deGennes, ChaikinLubensky, Martin},
but they are also consistent with the findings in 
the quantum-Hall smectic \cite{MacDonaldFisher}. As will become clear, they arise 
in the present context in a quite surprising way.

\begin{figure}[t]
\begin {center}
\includegraphics[width=0.15\textwidth]{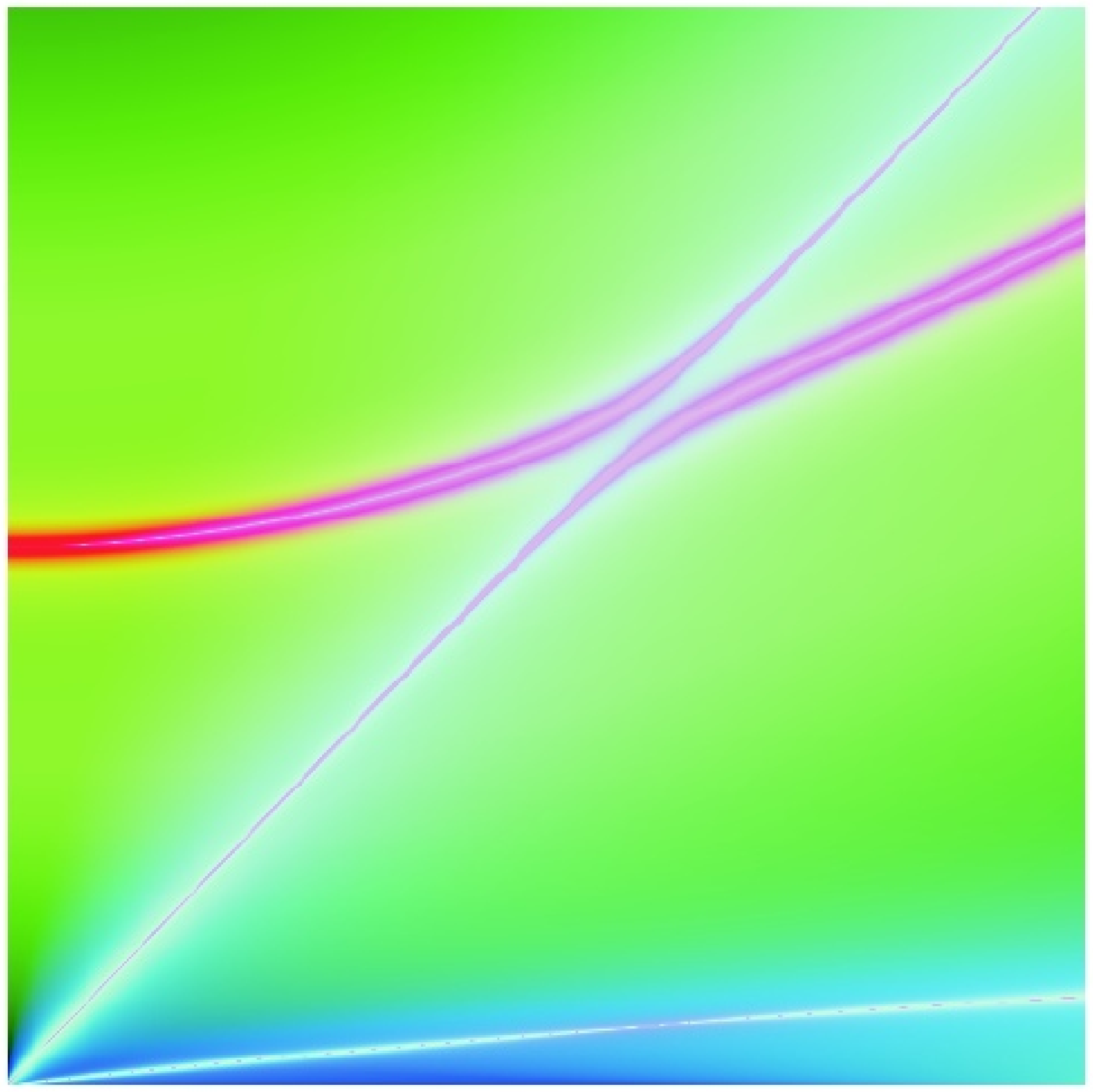}
\includegraphics[width=0.15\textwidth]{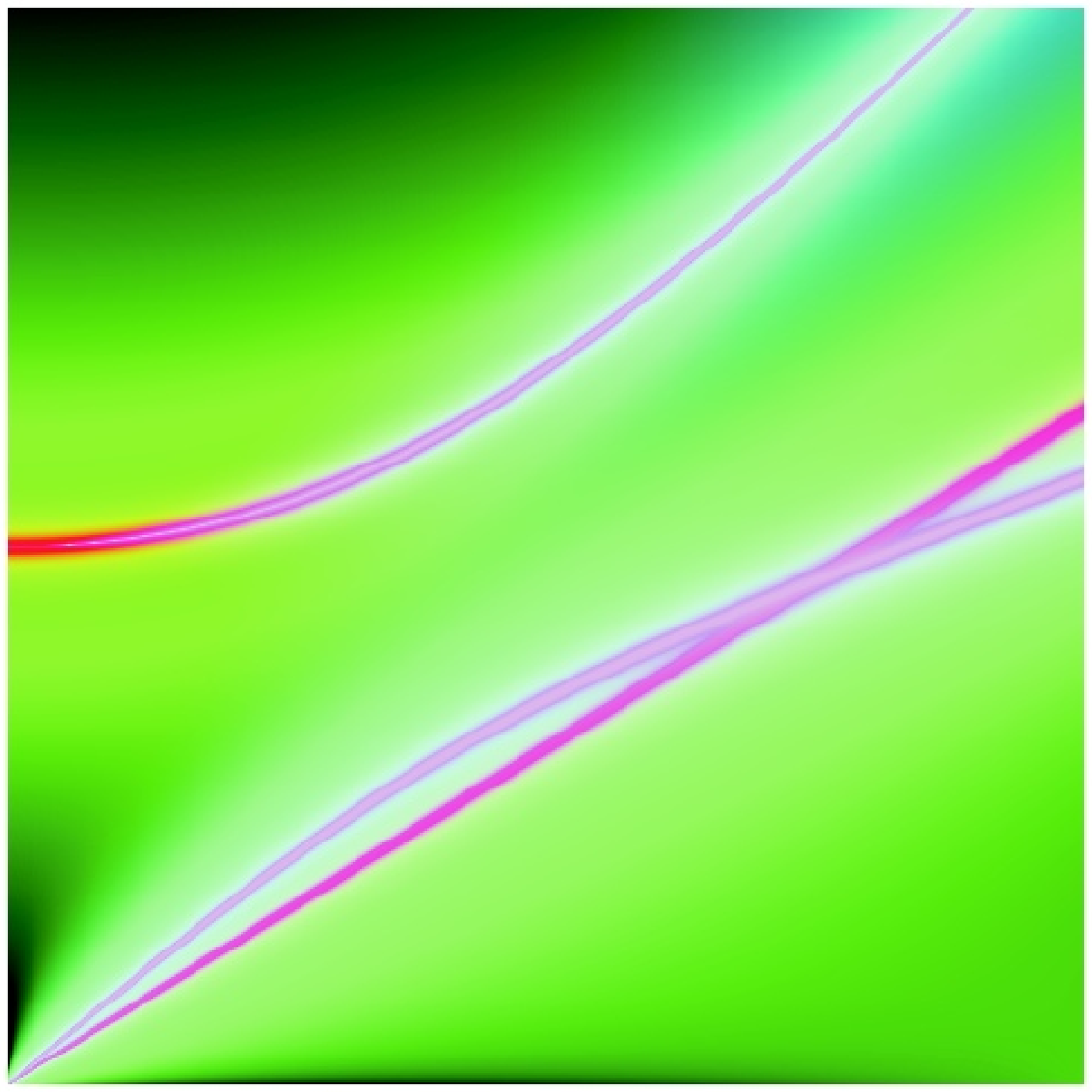}
\includegraphics[width=0.15\textwidth]{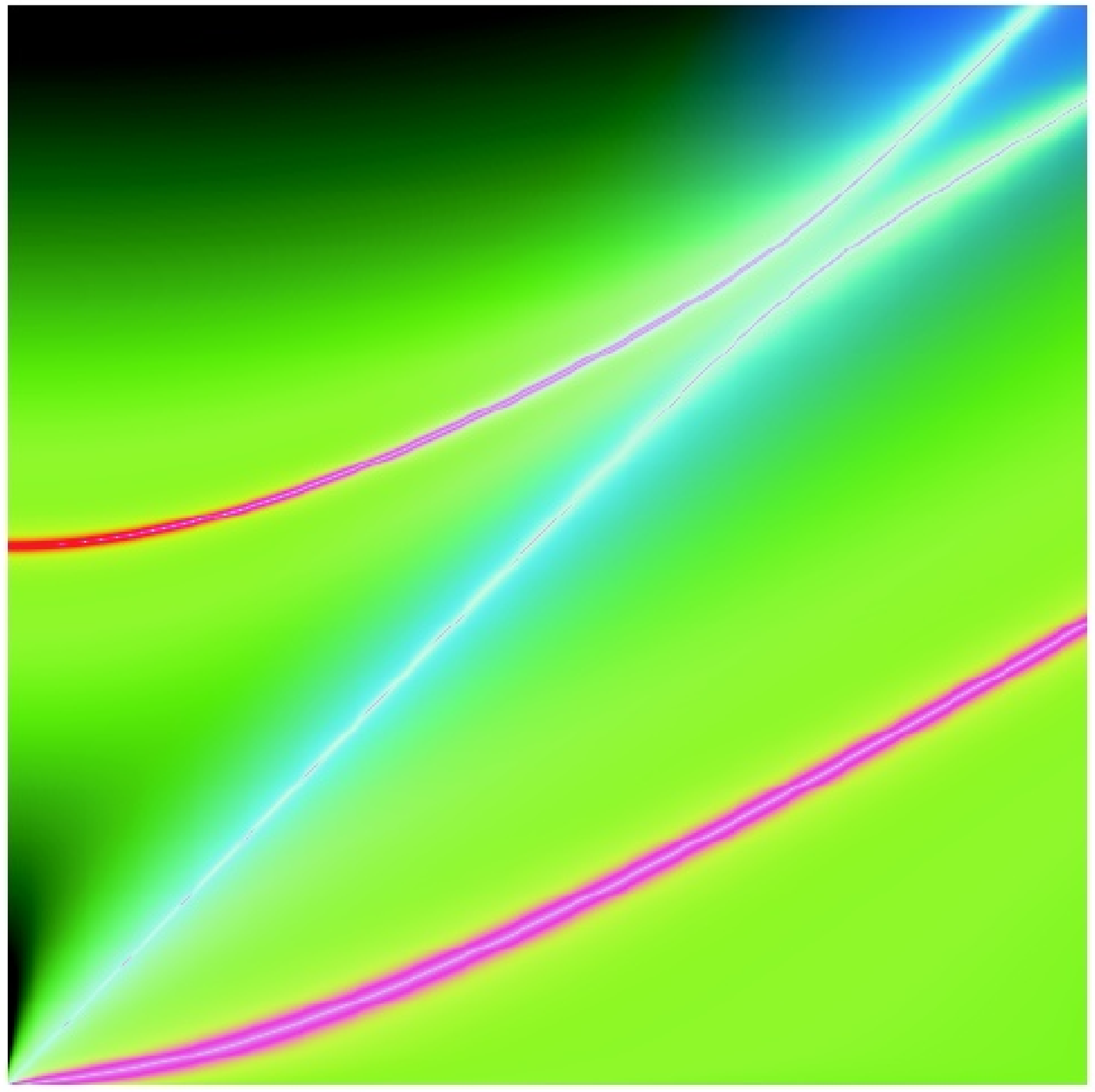}
\end {center}
\begin {center} a) \hspace {0.13\textwidth} b)  \hspace {0.13\textwidth} c) \end {center}
\caption{ (color). The long-wavelength  modes at  intermediate
values of the propagation angle $\eta$. (a) Close to the
`solid' direction ($\eta = 14 \pi / 30$) the longitudinal
`phonon' mode and massive shear photon of Fig.\ \ref {FigBig}a are coupled,
but in addition a massless mode has appeared with a velocity
and pole strength in the strain propagator increasing $\propto
\eta$. (b) Halfway ($\eta = \pi/4$), one finds two massless
modes with velocities $c_T$ and the compressional velocity
$c_{\kappa}$, as well as the massive shear photon. (c) Close
to the `liquid' direction ($\eta = \pi /30$), the quadratic mode
of Fig.\ \ref {FigBig}g acquires a linear dispersion at small momenta.}   
\label {Figeta}
\end{figure}

What happens when one deviates from either the liquid or solid direction? The
transversal and longitudinal modes now couple and at arbitrary angles 
a `longitudinal like' mode is present with a velocity
decreasing from $c_L$ to, amusingly, a purely compressional velocity
$c_{\kappa} = \sqrt{ \kappa / \rho}$ at $\pi/4$ (Fig.\ \ref {Figeta}b).
In the transversal-like response one finds always the dual Higgs photon.
However, deviating infinitisimally from the solid direction ($\eta = \pi/2$)
one finds immediately a new massless pole in the spectrum  with a velocity
$c_1 \approx c_T \eta \sqrt {2 (1+\nu)}$ demonstrating that shear
rigidity  becomes finite (Fig.\ \ref {Figeta}a). This mode
persists at all intermediate angles, with a velocity going through a maximum
at $\eta = \pi/4$. Upon approaching the liquid
direction ($\eta =0$, Fig.\ \ref {Figeta}c) it continues to be the only 
`transversal-like' massless mode which turns
into the quadratic mode (Fig.\ \ref {FigBig}g) precisely at $\eta =0$.

To the best of our knowledge, the massles excitations  amount  to
a novel enumeration of the Goldstone sector of the quantum-smectic state. 
The way we derived it might be surprising: it can be viewed
as the observable `shadow' of an underlying order which we
call the {\em dual shear superconductor}. It is about  a duality construction
which is a close sibbling of the well known vortex duality  in 
2+1D \cite {superfluid,XYdual,reldual}:
the quantum disordered superfluid (Mott-insulator) corresponds to
a bose condensate of vortices. Since the long range interactions between the 
vortices are  indistinguishable from electromagnetic interactions,
this dual condensate is a (gauged) superconductor. It is even a relativistic 
Higgs condensate: the particle and hole excitations
of the Mott insulator are identified with the `massive Goldstone'
and `longitudinal' photon of the relativistic  theory \cite {reldual}.
In dual elasticity the role of electromagnetic fields is taken by
the stress fields, hence the dual condensate of dislocations is
a (gauged) superconductor where (dual) shear stress becomes
short-ranged.

\begin{figure}[t]
\begin {center}
\includegraphics[width=0.46\textwidth]{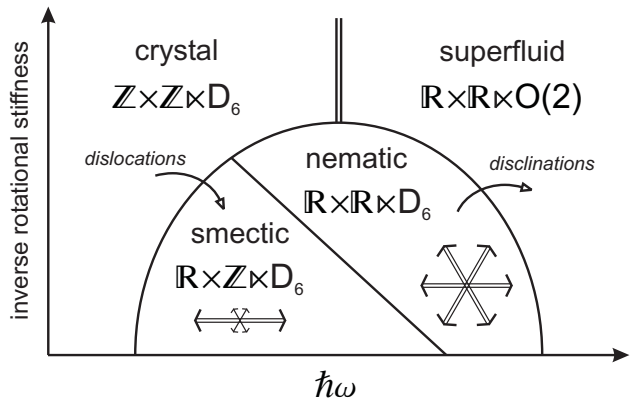}
\end {center}
\caption{The topology of the zero temperature 
phase diagram as suggested by
`quantum double Hopf symmetry breaking' \cite {BaisMathy}.
The amount of quantum disorder (i.e., coupling constant 
$\hbar \omega$) is represented on the horizontal axis; the vertical axis
carries the inverse rotational stiffness, a parameter whose smallness
prohibits the proliferation of disclinations \cite {thesis, Kleinert2}
that would lead to the isotropic superfluid phase.
The nematic or `quantum hexatic' state with the occupation
of all allowed Burgers vectors with equal probability while
the smectic corresponds in this topological scheme to a
preferential occupation one particular dislocation-antidislocation
pair.}\label {FigBais}
\end{figure}

Turning to the quantum crystal, the breaking of translations
implies shear rigidity and the translational topological charge
is the Burgers vector of the dislocation \cite{Friedel}.
Disclinations take 
care of rotations and in topological language a (quantum)
liquid crystal is a medium where dislocations have proliferated
while disclinations stay massive, as pointed out a long time ago
in the classical context by Nelson, Halperin and Young \cite{NHY}. 
This classic work deals with nematic (`hexatic') order and
how does one fit in smectics? In this regard we profit from a very 
recent theoretical advance. In the construction of the dislocation
condensates one has room to play with the vectorial topological
charges, and according to Bais and Mathy \cite {BaisMathy}, 
the usual recipy for spontaneous
symmetry breaking based on Lie groups does not suffice to classify defect condensates.
Instead, one needs Hopf symmetry, a larger mathematical
structure keeping track of symmetry and topology simultaneously,
with the `quantum double' Hopf symmetry 
being the one of relevance to liquid crystalline 
orders \cite {BaisMathy}. Their bottomline is illustrated in 
Fig.\ \ref {FigBais}: consider for simplicity the 
hexagonal crystal breaking  the symmetry of 2D Euclidean 
space ${\mathbb R}^2 \ltimes O(2)$ to ${\mathbb Z}^2 \ltimes D_6$.
The Burgers vectors can take values along any of the six directions in the hexagonal lattice. According to the quantum-double machinery, one possibility is that the 
Burgers vectors are populated with equal probability
along all six directions and the overall symmetry of this state is
${\mathbb R}^2 \ltimes D_6$: the hexatic nematic state,
where all translations are restored but rotations are broken.
However, it is also allowed to condense 
dislocation-antidislocation pairs {\em in one particular direction} and the symmetry of this state is ${\mathbb R} \times {\mathbb Z} \ltimes D_6$, which
is the symmetry of the smectic! Hence, smectic order is part
of the repertoire of topological melting, and at zero temperature
it should be located in between the crystal and the hexatic 
(Fig.\ \ref {FigBais}).
 
The above amounts to a symmetry classification and  next
is the construction of the theory describing the dynamics
in 2+1D.  Since dislocations are Abelian defects, this
follows the pattern of vortex duality resulting
in a dual gauge theory. Adapting the methods developed by 
Kleinert in the 1980's for the 3D classical 
case \cite {Kleinert2} we derived the 2+1D theory in an
earlier paper \cite{ZMN}, but this contains
some technical errors, which are serious enough to obscure
the physics: the neglect of the condensate dynamics and
a flawed gauge fix.  These are now repaired \cite{thesis},
and we will present here a  summary of this earlier
work, to discuss in detail the correct treatment of
the dual shear superconductor.  

Consider  a hexagonal crystal in 2+1D with
a long wavelength action coinciding with isotropic quantum-elasticity:
the usual theory in terms of the strains $\partial_a u^b$ ($u^b$'s
are the atomic displacements)
and an added kinetic energy $(\rho /2 ) (\partial_{\tau} u^a)^2$.
In terms of the stress field $\sigma^a_{\mu}$
($\mu = x,y, \tau$ and $a = x,y$), dual to the strains
$\partial_{\mu} u^a$, the Lagrangian reads in 2+1D Euclidean
space-time \cite{Kleinert2,ZMN},
\begin{equation}
  {\mathscr L}_{gauge}  = \tfrac 1 2 \sigma_\mu^a C_{\mu \nu ab}^{-1} \sigma_\nu^b
\label{stressaction}
\end{equation}
where $\sigma^a_{\tau}$ correspond to the momenta dual 
to the velocities $\partial_{\tau} u^a$, with inverse coupling constant 
$C_{\tau \tau a b}^{-1} = \delta^{ab} / \rho$. The conservation
of stress implies $\partial_{\mu} \sigma_{\mu}^a
= 0$ and this can be imposed in 2+1D by expressing
the stress-fields in terms of {\em stress gauge
fields} \cite{Kleinert2} $B^a_{\mu}$,
\begin{equation}
  \sigma^a_{\mu} = \epsilon_{\mu \nu \lambda} \partial_{\nu} B^a_{\lambda}
  \label{stressgauge}
\end{equation}
Hence, the theory describing acoustic phonons can be formulated
as a Maxwell theory  containing two `flavored' ($a = x,y$) 
$U(1)$ gauge fields --- these photons describe the capacity of the 
solid to carry elastic forces. Imposing the Ehrenfest 
constraint ($\sigma_x^y = \sigma_y^x$), 
one ends up with a Coulomb force between static sources and 
two propagating photons, coinciding with the longitudinal 
and transversal phonon of the 2+1D medium \cite{ZMN,thesis}. 
The sources of these stress gauge fields are the non-integrabilities 
associated with the translational part of the displacement 
fields \cite{Kleinert2,ZMN},
\begin{equation}
   {\mathscr L}_{int}  =  i B^a_{\mu} J^a_{\mu}, \; \; \;
  J^a_{\mu}  =  \epsilon_{\mu \nu \lambda} 
  \partial_{\nu} \partial_{\lambda} u^a.
  \label{dislcurrents}
\end{equation}
These  dislocation currents
can be factorized as $J_\mu^a = b^a {\mathscr J}_\mu$,
the world-lines of dislocations  with
Burgers vector $\vec{b} = ( b_x, b_y)$, implying
$B_\mu^a J_\mu^a \to (b^a B_\mu^a) {\mathscr J}_\mu$. This
structure is quite similar to vortex duality: the dislocations
are `charged' particles (like the vortices) minimally coupled
to gauge fields mediating the long range interactions, except 
that these fields are now stress gauge fields, while
the `Burgers' charge is vectorial. 

The kinematics of dislocations is unusual because of the
`glide constraint', the fact 
that dislocations only propagate  in  the direction of 
their Burgers vectors \cite{Friedel}, implying 
that the space-like components of the dislocation
currents form symmetric tensors 
$J^x_y - J^y_x = 0$ \cite{ZMN,glide}. 
A ramification is that dislocations only carry shear, and no compressional 
gauge charge. The effect is that 
only shear rigidity is destroyed by the dislocation condensate,
while it still carries  
sound \cite{ZMN,glide}. Glide plays an especially interesting 
role in the quantum smectic: although the dislocations form a 2+1D
condensate, the Burgers vectors are oriented in the liquid
direction and only in this direction `diamagnetic' currents 
occur,  screening the shear stress. This fact is largely
responsible for the dichotomy highlighted in Fig.\ \ref{FigBig}.

We overlooked in our previous work \cite {ZMN} the fact that 
the velocity  scale of the dislocation condensate coincides
with the transversal phonon velocity, playing the role of light
velocity: the dual shear superconductor is a relativistic
`Higgs phase'. This is implied by the effective Lorentz invariance
of the starting theory \cite{reldual}, where space and time
derivatives appear on equal footing
such that the phonon velocity also governs the dynamics of the defects \cite{Friedel}. 
Having collected all the required information, let us not turn to the order parameter
theory of the smectic dual shear superconductor \cite{thesis}.
Given the dislocation currents Eq.~(\ref{dislcurrents}), the tangle of dislocation
world lines can be described in terms of a single
\cite{ZMN} Ginzburg-Landau-Wilson (GLW) order
parameter field $\Psi$ as,
\be
  {\mathscr L}_{GLW} &=& 
  \tfrac 1 2 |(\partial_\mu - i b^a B_\mu^a) \Psi|^2 + V (\Psi) + {\mathscr L}_{gauge}. 
  \label {L_GLW}
\ee
With the usual potential term 
$V(\Psi) = \tfrac \lambda 2 (|\Psi|^2 - \Omega^2 / \mu)^2$ and the Maxwell-like 
term ${\mathscr L}_{gauge}$ defined by Eq. (\ref{stressaction}). In the dislocation
condensate  amplitude has condensed, $|\Psi_0| = \Omega / \sqrt \mu$, and by integrating out the dislocation phase field the
remaining Higgs term can be written in  
gauge invariant form as \cite{Kleza},
\be
  {\mathscr L}_{Higgs, bare} = \tfrac 1 2 \frac {\Omega^2}{\mu}
  \frac {(b^a \sigma_\mu^a) (b^b \sigma_\mu^b)}{\partial_\mu^2}, \label {L_Higgs_bare}
\ee
with $\partial_\mu^2 = \tfrac 1{c_T^2} \partial_\tau^2 +
\partial_i^2$, using the condensate velocity $c_T$. The glide constraint
has still to be imposed and this is
straightforwardly accomplished by adding a Lagrange multiplier to the action,
\be
  {\mathscr L}_{glide} = i \lambda \varepsilon_{\tau \mu a} J_\mu^a,
\ee
which implies that the gauge fields in Eq.~(\ref {L_GLW})
should acquire an additional piece $B_\mu^a \to B_\mu^a +
\lambda \varepsilon_{\tau \mu a}$, or equivalently that
the dual stresses turn into $\sigma_\mu^a \to \sigma_\mu^a
+ \delta_{\mu a} \partial_\tau \lambda - \delta_{\mu \tau} \partial_a \lambda$.
Integrating out the multiplier field $\lambda$ we obtain a Higgs term respecting
the glide constraint
\be
  {\mathscr L}_{Higgs} = \tfrac 1 2 \frac {\Omega^2}{\mu}
  \frac {(\sigma_H)^2}{(\partial_\mu^{(\eta)})^2}. \label {L_Higgs_glide}
\ee
with the dislocation second sound governed by
\be
  (\partial_\mu^{(\eta)})^2 = \tfrac {1}{c_T^2} \partial_\tau^2 +
  b^a \partial_a^2 \leadsto \tfrac 1 {c_T^2}\omega_n^2 + q^2 \cos^2 \eta.
\ee
showing that the condensate only propagates along the direction of the
Burgers vectors, introducing also the angle $\eta$ keeping track of  the
direction of the liquid direction relative to the wave-vector. In addition, 
only the following stress component appears in the Higgs term,
\be
  \sigma_H = \epsilon_{ac} b^a b^b \sigma_c^b. \label {sigmaH}
\ee
which becomes in the  quantum smectic   $\sigma_H = \sigma^x_y$,
taking the liquid direction along $\hat{x}$ such that
$b^b b^c  = \delta_{b,x}\delta_{c,x}$. The remainder is 
a lengthy but straightforward exercise:
Eqs.~(\ref{stressaction},\ref{L_Higgs_glide}) are expressed
in the stress gauge fields via Eq.~(\ref{stressgauge}), followed
by imposing gauge transversality (using
a Lorentz gauge  $\tfrac 1{c_T^2} \partial_\tau B_\tau^a + \partial_i B_i^a = 0$)
and the Ehrenfest constraint. Finally,  stress-strain
relations are used to express the strain (phonon)
propagators $\langle \langle \partial u | \partial u \rangle \rangle$
in terms of stress photon propagators \cite{ZMN,reldual} 
and the spectral functions of the former
are shown in Figs.\ \ref {FigBig} and \ref {Figeta}.

We are now in the position to discuss some of the deeper
issues related to the mode spectrum, Figs.\ \ref {FigBig} and \ref {Figeta}.
There are three propagating modes: the two (gauge-transversal) 
stress photons of the crystal plus the single `longitudinal'
(or condensate) photon. For arbitrary $\eta$'s the mode
couplings conspire to produce the two massless `phonons'
and the massive shear photon of  Fig.\ \ref {Figeta}.
The longitudinal photons decouple completely from the condensate
in the `liquid' and `crystal' directions of Fig.\ \ref {FigBig},
but this is different in the  transversal sectors: in the solid direction
only the Higgsed transversal `phonon' shows up, while
the `longitudinal photon' cannot be excited because of the
one dimensional (glide) nature of the shear screening currents. 
In the liquid direction this is different: the massive mode is the
massive second sound of the condensate (longitudinal photon), while the quadratic 
mode is like a Higgsed gauge particle where the `Higgs mechanism failed at the
last moment'!  

If quantum smectics exist, are its modes observable? In
Figs.~\ref {FigBig} and \ref {Figeta} we show 
phonon propagators, but there is no obvious `atomistic' candidate quantum
smectic. In electronic systems, the smectic modes can only be
detected by their electromagnetic response, a story by itself
which we will discuss elsewhere \cite{natphys}. 
What about DC properties? The dual shear superconductor 
deals with dimensionality in a strange way: the dislocations 
form an ordered 2+1D Higgs condensate, carrying only one
dimensional shear screening currents because of the glide 
principle. On `our side' of the duality this has the strange
consequence that it would actually be a superfluid but only
precisely along the liquid direction \cite{ZMN, thesis}. This is, however,
a pathology of the (implicit) assumption governing the field
theory that all length scales are large compared to the lattice
constant. This is equivalent to the
assumption that the single-particle (interstitial and vacancy)
defects can be ignored which is not the case for any finite
lattice constant \cite{ZMN}. A  real life Bose quantum smectic should therefore behave 
as an anistropic two dimensional superfluid.

We acknowledge helpful discussions with E.\ Demler,
S.I.\ Mukhin, and Z.\ Nussinov. This work was supported 
by the Netherlands foundation for fundamental research 
of Matter (FOM).

\bibliographystyle{apsrev}

\end {document}